\documentclass[style=bcnpostgrad,entry]{proceedings}
\renewcommand{\conferencetitle}{Barcelona Postgrad Encounters on Fundamental Physics}
\usepackage[utf8]{inputenc}

\usepackage{latexsym}
\usepackage{epsfig}
\usepackage{amssymb}

\begin{document}

\author[1,a]{Tirthabir Biswas}
\author[2,b]{\underline{Tomi Koivisto}}
\author[3,c]{Anupam Mazumdar}

\email[1]{tirthabir@gmail.com}
\email[2]{tomik@astro.uio.no}
\email[3]{a.mazumdar@lancaster.ac.uk}

\affiliation[a]{Physics Department, Loyola University, Campus Box 92, New Orleans, LA 70118}
\affiliation[b]{Institute of Theoretical Astrophysics, University of Oslo, N-0315 Oslo, Norway}
\affiliation[c]{Consortium for Fundamental Physics, Lancaster University, Lancaster, LA1 4YB, UK}



\title{Nonlocal theories of gravity: the flat space propagator} 

\abstract{It was recently found that there are classes of nonlocal gravity theories that are free of ghosts and singularities in their Newtonian limit [PRL, 108 (2012), 031101]. 
In these proceedings, a detailed and pedagogical derivation of a main result, the flat space propagator for an arbitrary covariant metric theory of gravitation, is presented. The result is applied to analyse  f(R) models, Gauss-Bonnet
theory, Weyl-squared gravity and the potentially asymptotically free nonlocal theories.} 

\maketitle

\section{Introduction}

General Relativity (GR) predicts singularities and doesn't straightforwardly yield to quantisation. On the other hand, attempts to modify the theory are restricted by, besides phenomenological viability, theoretical consistency. By straying away from the Einstein-Hilbert action for the metric of space-time, one easily invites ghosts into the theory. While some specific higher derivative theories may be free of ghosts, they are not renormalisable, and vice versa  \cite{Stelle:1976gc}.

However, nonlocal theories, featuring an infinite number of derivatives, might provide a way around this \cite{Biswas:2011ar,Modesto:2011kw}. The ultraviolet singularities may then be smoothened out without introducing ghosts and while recovering GR predictions at small curvatures. 
Indeed, the promising attempts at quantum gravity such as string theory and loop quantum gravity exhibit nonlocality at some fundamental level. Phenomenologically nonlocal gravity has been recently applied to model pre-big bang cosmology \cite{Biswas:2005qr,Biswas:2010zk}, inflation \cite{Capozziello:2008gu,Biswas:2012bp}, screening mechanisms  \cite{ArkaniHamed:2002fu, Zhang:2011uv}, dark energy \cite{Deser:2007jk,Koivisto:2008xfa}, structure formation \cite{Koivisto:2008dh,Park:2012cr} and dark matter \cite{Soussa:2003vv,Blome:2010xn}.  Theoretical studies have considered renormalisability \cite{Moffat:2010bh,Modesto:2012za} and black holes \cite{Modesto:2010uh,Nicolini:2012eu}.   

In these proceedings, we will derive the flat space propagator for the most general metric theory of gravity, presented in \cite{Biswas:2011ar}. In section \ref{qa} we write down the action and reduce it to a tractable form in the relevant Minkowski limit. Section \ref{pr} then introduces the formalism and the method to obtain the propagator. The result is applied in Section \ref{sc} for an analysis of several special cases of interest.

\section{The most General Quadratic Action}
\label{qa}

To understand both the asymptotic behavior and the ghost free condition, what is relevant is the quadratic action of gravity. In other words if we look at fluctuations around the Minkowski background
\begin{equation}
g_{\mu\nu}=\eta_{\mu\nu}+h_{\mu\nu}\ ,
\end{equation}
then all we need to worry about are the terms that are quadratic in $h_{\mu\nu}$ in the action. Now, since in the Minkowski background $R_{\mu\nu\lambda\sigma}$ vanishes, everytime there is a Riemann tensor in the action it contributes an ${\cal O}(h)$ term in the action. Thereby we only need to analyse  terms in the action that are products of at most  two curvature terms, and the most general form for the action is given by
\begin{equation}
S_q=\int d^4x\sqrt{-g}R_{\mu_1\nu_1\lambda_1\sigma_1}{\cal D}_{\mu_2\nu_2\lambda_2\sigma_2}^{\mu_1\nu_1\lambda_1\sigma_1}R^{\mu_2\nu_2\lambda_2\sigma_2}\,,
\end{equation}
where ${\cal D}$ is a differential operator containing covariant derivatives and $\eta_{\mu\nu}$. We note that if there is a differential operator acting on the left Riemann tensor as well, one can always recast that into the above form using integration by parts.

Since the operator ${\cal D}$ can only have covariant derivatives and the Minkowski metric, one can actually write down the most general action $S_q$ explicitly:
\begin{eqnarray} \label{sq}
S_q&=&\int d^4x\sqrt{-g}\Big[R F_1(\Box)R+R F_2(\Box)\nabla_{\mu}\nabla_{\mu}R^{\mu\nu}+R_{\mu\nu} F_3(\Box)R^{\mu\nu}+R_{\mu}^{\nu} F_4(\Box)\nabla_{\nu}\nabla_{\lambda}R^{\mu\lambda}\nonumber \\ 
&+&R^{\lambda\sigma} F_5(\Box)\nabla_{\mu}\nabla_{\sigma}\nabla_{\nu}\nabla_{\lambda}R^{\mu\nu}+R F_6(\Box)\nabla_{\mu}\nabla_{\nu}\nabla_{\lambda}\nabla_{\sigma}R^{\mu\nu\lambda\sigma}+R_{\mu\lambda} F_7(\Box)\nabla_{\nu}\nabla_{\sigma}R^{\mu\nu\lambda\sigma}\nonumber \\ 
&+&R_{\lambda}^{\rho} F_8(\Box)\nabla_{\mu}\nabla_{\sigma}\nabla_{\nu}\nabla_{\rho}R^{\mu\nu\lambda\sigma}+R^{\mu_1\nu_1} F_9(\Box)\nabla_{\mu_1}\nabla_{\nu_1}\nabla_{\mu}\nabla_{\nu}\nabla_{\lambda}\nabla_{\sigma}R^{\mu\nu\lambda\sigma} \nonumber \\ 
& + & R_{\mu\nu\lambda\sigma} F_{10}(\Box)R^{\mu\nu\lambda\sigma} + R_{\mu\nu\lambda}^{\rho} F_{11}(\Box)\nabla_{\rho}\nabla_{\sigma}R^{\mu\nu\lambda\sigma} \nonumber \\ 
& + & R_{\mu\rho_1 \nu\sigma_1} F_{12}(\Box)\nabla^{\rho_1}\nabla^{\sigma_1}\nabla_{\rho}\nabla_{\sigma}R^{\mu\rho\nu\sigma} +
R_{\mu}^{\nu_1\rho_1\sigma_1}F_{13}(\Box)\nabla_{\rho_1}\nabla_{\sigma_1}\nabla_{\nu_1}\nabla_{\nu}\nabla_{\rho}\nabla_{\sigma}R^{\mu\nu\lambda\sigma} \nonumber \\ 
& + & R^{\mu_1\nu_1\rho_1\sigma_1} F_{14}(\Box)\nabla_{\rho_1}\nabla_{\sigma_1}
\nabla_{\nu_1}\nabla_{\mu_1}\nabla_{\mu}\nabla_{\nu}\nabla_{\rho}\nabla_{\sigma}R^{\mu\nu\lambda\sigma}\Big]\,.
\end{eqnarray}
Using the antisymmetric properties of the Riemann tensor,
\begin{equation}
R_{(\mu\nu)\rho\sigma} = R_{\mu\nu(\rho\sigma)} = 0\,, \label{antis}
\end{equation}
and the Jacobi identity
\begin{equation}
\nabla_\alpha R_{\mu\nu\beta\gamma} + \nabla_\gamma R_{\mu\nu\beta\alpha} + \nabla_\beta R_{\mu\nu\gamma\alpha}\,, 
\end{equation}
one finds after patient index manipulation that the above action reduces to
\begin{eqnarray} \label{sq2}
S_q&=&\int d^4x\sqrt{-g}\Big[R F_1(\Box)R + R_{\mu\nu} F_3(\Box)R^{\mu\nu}
+ R F_6(\Box)\nabla_{\mu}\nabla_{\nu}\nabla_{\lambda}\nabla_{\sigma}R^{\mu\nu\lambda\sigma} \nonumber \\
& + & R_{\mu\nu\lambda\sigma} F_{10}(\Box)R^{\mu\nu\lambda\sigma}  +
R_{\mu}^{\nu_1\rho_1\sigma_1}F_{13}(\Box)\nabla_{\rho_1}\nabla_{\sigma_1}\nabla_{\nu_1}\nabla_{\nu}\nabla_{\rho}\nabla_{\sigma}R^{\mu\nu\lambda\sigma} \nonumber \\ 
& + & R^{\mu_1\nu_1\rho_1\sigma_1} F_{14}(\Box)\nabla_{\rho_1}\nabla_{\sigma_1}
\nabla_{\nu_1}\nabla_{\mu_1}\nabla_{\mu}\nabla_{\nu}\nabla_{\rho}\nabla_{\sigma}R^{\mu\nu\lambda\sigma}\Big]\,.
\end{eqnarray}
So we got rid of 8 of the 14 terms already in a curved background.


\subsection{Linearised Action}
Our next task is to obtain the quadratic (in $h_{\mu\nu}$) free part of the above action. A very important simplification occurs when we realise that the two $h$-dependent terms must come from the two curvature terms present. 
In other words the covariant derivatives must take on the Minkowski values, and we can commute them freely. We then observe  that the $F_6$, $F_{13}$ and $F_{14}$ terms in the action (\ref{sq2}) become irrelevant in this limit due to the symmetry of the derivative operations contracting the antisymmetric index pairs of the Riemann tensor (\ref{antis}). The linearised action contains in the end only
\begin{eqnarray} \label{f_action}
S_q&=&\int d^4x\left[R F_1(\Box)R
+R_{\mu\nu} F_3(\Box)R^{\mu\nu}
+R_{\mu\nu\lambda\sigma} F_{10}(\Box)R^{\mu\nu\lambda\sigma}\right]\,.
\end{eqnarray}
Furthermore, below it will become clear that there are essentially only two free functions that determine the properties of the theory in this limit.

Now our next task is to substitute the linearised expressions of the curvatures in terms of $h_{\mu\nu}$:
\begin{eqnarray}
R_{\mu\nu\lambda\sigma} & = & \frac{1}{2}(\partial_{[\lambda}\partial_{\nu}h_{\mu\sigma]}-\partial_{[\lambda}\partial_{\mu}h_{\nu\sigma]})\,, \nonumber \\ 
R_{\mu\nu}& = & \frac{1}{2}(\partial_{\sigma}\partial_{(\nu}h_{\mu)}^{\sigma}-\partial_{\nu}\partial_{\mu}h-\Box h_{\mu\nu})\,, \quad 
R = \partial_{\nu}\partial_{\mu}h^{\mu\nu}-\Box h\,. \nonumber
\end{eqnarray}
As is obvious, many of the terms simplify and combine. By considering all possible consistent contractions, one can deduce that all the terms eventually have to produce an action of the following form:
\begin{alignat}{5}
S_q = -\int d^4x\Big[
\frac{1}{2}h_{\mu\nu} \Box a(\Box)h^{\mu\nu}
& +  h_{\mu}^{\sigma} b(\Box)\partial_{\sigma}\partial_{\nu}h^{\mu\nu}
+h c(\Box)\partial_{\mu}\partial_{\nu}h^{\mu\nu} \nonumber \\
& +  \frac{1}{2}h\Box d(\Box)h
+h^{\lambda\sigma} \frac{f(\Box)}{\Box}\partial_{\sigma}\partial_{\lambda}\partial_{\mu}\partial_{\nu}h^{\mu\nu}\Big]\,,
\label{lin_act}
\end{alignat}
where we have defined the functions $a(\Box)$, $b(\Box)$, $c(\Box)$ and $d(\Box)$ in such a way that they 
reduce in the appropriate limit to the constants $a$, $b$, $c$ and $d$ used by van Nieuwenhuizen \cite{VanNieuwenhuizen:1973fi}. The function $f(\Box)$ appears 
only in higher order theories. 

We will now compute all of the terms in the original action (\ref{f_action}) individually. The first piece gives
\begin{equation}
R F_1(\Box)R = h F_1 \Box^2 h + h^{\lambda\sigma} F_1 \partial_{\sigma}\partial_{\lambda}\partial_{\mu}\partial_{\nu}h^{\mu\nu}
- h F_1 \Box \partial_\mu \partial_\nu h^{\mu\nu}
- h^{\mu\nu} F_1\Box \partial_\mu \partial_\nu h\,. \nonumber
\end{equation}
The third and fourth terms in this case can be combined as follows.  Ignoring surface terms it is always possible to commute through the local $F(\Box)$ terms\footnote{For non-polynomial terms that is not clear.}, and we get
\begin{equation}
R F_1(\Box)R = F_1 (\Box) \left[ h  \Box^2 h + h^{\lambda\sigma}  \partial_{\sigma}\partial_{\lambda}\partial_{\mu}\partial_{\nu}h^{\mu\nu}
- 2 h  \Box \partial_\mu \partial_\nu h^{\mu\nu} \right]\,. \nonumber
\end{equation}
For the two other relevant terms in the (\ref{f_action}) we obtain in the similar way: 
\begin{eqnarray}
R_{\mu\nu} F_3 (\Box) R^{\mu\nu} & = & F_3(\Box) \Big[
{\frac{1}{4}} h \Box^2 h + \frac{1}{4} h_{\mu\nu}  \Box^2 h^{\mu\nu}
-\frac{1}{2} h_{\mu}^{\sigma} \Box \partial_{\sigma}\partial_{\nu}h^{\mu\nu} \nonumber \\ 
& - & \frac{1}{2} h \Box \partial_{\mu} \partial_{\nu} h^{\mu\nu} 
 +  \frac{1}{2} h^{\lambda\sigma}  \partial_{\sigma}\partial_{\lambda}\partial_{\mu}\partial_{\nu}h^{\mu\nu}\Big]\,; \nonumber \\ 
R_{\mu\nu\lambda\sigma} F_{10}(\Box)R^{\mu\nu\lambda\sigma} & = &
F_{10}(\Box) \left[ h_{\mu \nu}  \Box^2 h^{\mu \nu}
+ h^{\lambda\sigma} \partial_{\sigma}\partial_{\lambda}\partial_{\mu}\partial_{\nu}h^{\mu\nu}
- 2 h_{\mu}^{\sigma}  \Box \partial_{\sigma}\partial_{\nu}h^{\mu\nu}\right]\,.  \nonumber
\end{eqnarray}
It remains to relate these terms to the five combinations appearing in action ({\ref{lin_act}).
\subsection{The coefficients in terms of $F_i(\Box)$}
In the above section we have calculated the contribution of the higher derivative modifications (\ref{f_action}) to the action (\ref{lin_act}). We also need to include the contribution from the Einstein-Hilbert term, so that the full action we consider is
\begin{equation}
S=\int d^4x\sqrt{-g}R + S_q\,.
\end{equation}
We then eventually obtain
%
%
\begin{alignat}{5}
a(\Box) &=   1-\frac{1}{2} F_3 (\Box) \Box 
- 2 F_{10} (\Box) \Box\,,
\notag \\
b(\Box) &=  -1+\frac{1}{2} F_3 (\Box) \Box  + 2 F_{10} (\Box) \Box\,,
\notag \\
c(\Box) &= 1+ 2 F_1(\Box) \Box + \frac{1}{2} F_3 (\Box) \Box\,,
\notag \\
d(\Box) &= -1-2F_1(\Box) \Box- \frac{1}{2} F_3 (\Box) \Box\,,
\notag \\
f(\Box)  &=   -2F_1 (\Box) \Box  -  F_3 (\Box)\Box- 2 F_{10}(\Box) \Box\,. \label{coef}
\end{alignat}

From the above expressions we observe the following interesting relations\footnote{An immediate observation we can make is that one cannot construct the Fierz-Pauli term, for which $a=-d\sim m^2$ and $b=c=0$ from an action like (\ref{sq}). Massive gravity is not among the metric theories we consider here, but inherently bimetric.} 
\begin{eqnarray}
a +b = 0 \,, \quad 
c +d = 0\,, \quad
b+c + f = 0 \,,
\label{Grelations}
\end{eqnarray}
so that we are really left with two independent arbitrary functions.
This can be understood as a consequence of the Bianchi identities, as will be shortly clarified.

\subsection{Field Equations \& Bianchi identities}

The field equations can be derived straightforwardly by varying the action (\ref{lin_act}): 
\begin{eqnarray}
a(\Box) \Box h_{\mu\nu}  & + &  b(\Box)\partial_{\sigma}\partial_{(\nu}h_{\mu)}^{\sigma}
 + c(\Box)(\eta_{\mu\nu}\partial_{\rho}\partial_{\sigma}h^{\rho\sigma} +\partial_{\mu}\partial_{\nu}h) \nonumber \\
 +   \eta_{\mu\nu}d(\Box)\Box h
& + &  f(\Box) \Box^{-1} \partial_{\sigma}\partial_{\lambda}\partial_{\mu}\partial_{\nu}h^{\lambda\sigma}  =  -\kappa\tau_{\mu\nu}\,.
\label{linearized-eqn}
\end{eqnarray}
The matter side is conserved by the stress energy conservation and the geometric part because of the generalised Bianchi identities \cite{Koivisto:2005yk} due to diffeomorphism invariance. Thus
\begin{eqnarray}
-\kappa\tau\nabla_\mu \tau^{\mu}_\nu = 0 =  (c+d)\Box \partial_\nu h +  (a+b)\Box h^\mu_{\nu,\mu} + (b+c+f )h^{\alpha\beta}_{,\alpha\beta\nu}\,.
\end{eqnarray}
It is then clear why (\ref{Grelations}) had to hold.

The above field equations can be written in the form
\begin{equation} \label{inverse}
\Pi_{\mu\nu}^{-1}{}^{\lambda\sigma}h_{\lambda\sigma}=\kappa\tau_{\mu\nu}\,,
\end{equation}
where $\Pi_{\mu\nu}^{-1}{}^{\lambda\sigma}$ is the inverse propagator. To compute the propagator, we need to learn to deal with spin projector operators.

\section{Propagators}
\label{pr}

\label{p_app}

In this section we are going to derive the propagators for the field equations (\ref{linearized-eqn}). The basic algorithm is as follows~\cite{VanNieuwenhuizen:1973fi}: First, we express the field equations in the form ({\ref{inverse}),
where the inverse propagator $\Pi^{-1}$ is expressed in terms of the six operators, ${\cal P}_i$ (to be specified shortly):
\begin{equation}
\Pi^{-1}=\sum_{i=1}^6 C_i{\cal P}_i
\label{operators}
\end{equation}
In the momentum space description of the coefficients $C_i$'s are scalars which can  only depend on $k^2$. Finding the suitable operators, it is possible to decompose the field equations into a decoupled set of equations of motion for the relevant degrees of freedom. These are then conveniently invertible.

\subsection{Spin projector operators}  
  
 Let us introduce
\begin{eqnarray}
{\cal P}^2&=&\frac12(\theta_{\mu\rho}\theta_{\nu\sigma}+\theta_{\mu\sigma}\theta_{\nu\rho})-\frac13\theta_{\mu\nu}\theta_{\rho\sigma}\,,
\nonumber\\
{\cal P}^1&=&\frac12(\theta_{\mu\rho}\omega_{\nu\sigma}+\theta_{\mu\sigma}\omega_{\nu\rho}+\theta_{\nu\rho}\omega_{\mu\sigma}+\theta_{\nu\sigma}\omega_{\mu\rho})\,,
\nonumber\\
{\cal P}^0_s&=&\frac13\theta_{\mu\nu}\theta_{\rho\sigma}\,, \quad
{\cal P}^0_w=\omega_{\mu\nu}\omega_{\rho\sigma}\,,
\nonumber\\
{\cal P}^0_{sw}&=&\frac1{\sqrt3}\theta_{\mu\nu}\omega_{\rho\sigma}\,, \quad
{\cal P}^0_{ws}=\frac1{\sqrt3}\omega_{\mu\nu}\theta_{\rho\sigma}\,,
\end{eqnarray}
where the transversal and longitudinal projectors in the momentum space are respectively
\begin{equation}
\theta_{\mu\nu}=\eta_{\mu\nu}-\frac{k_\mu k_\nu}{k^2},\qquad \omega_{\mu\nu}=\frac{k_\mu k_\nu}{k^2}.
\nonumber
\end{equation}
 Note that the operators ${\cal P}^i$ are in fact 4-rank tensors, ${\cal P}^i_{\mu\nu\rho\sigma}$, but we have suppressed the index notation here. 
 
 Out of the six operators four of them,  $\{{\cal P}^2,{\cal P}^1,{\cal P}_s^0,{\cal P}_w^0\}$, form a complete set of projection operators:
\begin{equation}
{\cal P}^{i}_{a}{\cal P}^{j}_{b}=\delta^{ij}\delta_{ab}{\cal P}^{i}_{a}
\quad
{\mbox and}
\quad
{\cal P}^2+{\cal P}^1+{\cal P}_s^0+{\cal P}_w^0=1\,,
\label{completeness}
\end{equation}
as one can easily verify.
These projection operators together  represent the six field degrees of freedom, the additional four fields in a symmetric tensor field, as usual, being gauge degrees of freedom.
${\cal P}^2$ and ${\cal P}^1$ represent transverse and traceless spin-2 and spin-1 degrees, accounting for four field degrees of freedom, while ${\cal P}_s^0$, ${\cal P}_w^0$ represent the spin-0 scalar multiplets. In addition to the above four spin operators we have ${\cal P}_{sw}^0$ and ${\cal P}_{ws}^0$ which can potentially mix the two scalar multiplets. In particular, we have that
\begin{equation}
{\cal P}^0_{ij}{\cal P}^0_k=\delta_{jk}{\cal P}^0_{ij}\,, \quad {\cal P}^0_{ij}{\cal P}^0_{kl} = \delta_{il}\delta_{jk}{\cal P}^0_k\,, \quad {\cal P}^0_k {\cal P}^0_{ij} = \delta_{ik}{\cal P}^0_{ij}\,,
\end{equation}
as one may again easily check.

From  (\ref{operators}) and (\ref{completeness}) it trivially follows that we can write (\ref{inverse}) as
\begin{equation}
\sum_{i=1}^6 C_i{\cal P}_i h=\kappa ({\cal P}^2+{\cal P}^1+{\cal P}_s^0+{\cal P}_w^0)\tau\,.
\label{operator-form}
\end{equation}
By multiplying with the different projector operators on either side of the equation we can now obtain the decoupled field equations for the different spin multiplets.  

\subsection{Inverting the field equations}

Having outlined the algorithm for finding the propagators, let us now proceed to obtain them in our model specified by the action (\ref{lin_act}). We need to express all the operators in (\ref{linearized-eqn}) in terms of the operators ${\cal P}^i$. Let us start with
\begin{eqnarray}
\eta_{\mu\nu}d(\Box)h&\rightarrow&  d(-k^2)\eta_{\mu\nu}\eta^{\rho\sigma}h_{\rho\sigma}=d(-k^2)(\theta_{\mu\nu}+\omega_{\mu\nu})(\theta^{\rho\sigma}+\omega^{\rho\sigma})h_{\rho\sigma} \nonumber \\
&=& d(-k^2)(\theta_{\mu\nu}\theta^{\rho\sigma}+\omega_{\mu\nu}\theta^{\rho\sigma}+\theta_{\mu\nu}\omega^{\rho\sigma}+\omega_{\mu\nu}\omega^{\rho\sigma})h_{\rho\sigma}\nonumber \\
&=& d(-k^2)[3{\cal P}_s^0+{\cal P}_w^0+{\sqrt{3}}({\cal P}_{sw}^0+{\cal P}_{ws}^0)]h\,. \nonumber
\end{eqnarray}
One can continue in an analogous fashion to obtain the projector decomposition of all the operators appearing in the field equations (\ref{linearized-eqn}). For the first three terms we then obtain
\begin{eqnarray}
a(\Box)h_{\mu\nu} & \rightarrow & a(-k^2)\left[{\cal P}^2+{\cal P}^1+{\cal P}^0_s+{\cal P}^0_w\right] h\,, \nonumber
\\
b(\Box)\partial_\sigma \partial_{(\nu}h^\sigma_{\mu)} & \rightarrow & -b(-k^2)k^2\left[ {\cal P}^1 + 2{\cal P}^0_w\right] h\,, \nonumber
\\
c(\Box)(\eta_{\mu\nu}\partial_\rho \partial_\sigma h^{\rho\sigma} + \partial_\mu \partial_\nu h )  & \rightarrow &
-c(-k^2)k^2 \left[ 2{\cal P}^0_w + \sqrt{3}\left( {\cal P}^0_{sw}+{\cal P}^0_{ws}\right) \right]  h\,. \nonumber
\end{eqnarray}
While all the above operators appear in two derivative generalisations of gravity and were discussed in~\cite{VanNieuwenhuizen:1973fi}, the $f$ term in (\ref{linearized-eqn}) is specific to higher derivative theories. Its decomposition is rather simple
\begin{eqnarray}
f(\Box)\partial^{\sigma}\partial^{\rho}\partial_{\mu}\partial_{\nu}h_{\rho\sigma}\rightarrow f(-k^2)k^{\sigma}k^{\rho}k_{\mu}k_{\nu}h_{\rho\sigma}=f(-k^2)k^4\omega_{\mu\nu}\omega^{\rho\sigma}=f(-k^2)k^4{\cal P}_w^0\,.\nonumber
\end{eqnarray}

We are now ready to write down the projected field equations, and the corresponding propagators. By acting with ${\cal P}^2$ on (\ref{operator-form}) we find
\begin{equation}
ak^2{\cal P}^2h=\kappa {\cal P}^2\tau\Rightarrow {\cal P}^2h=\kappa\left( {{\cal P}^2\over ak^2}\right)\tau\,.
\end{equation}
Similarly, acting with ${\cal P}^1$, one finds
\begin{equation}
(a+b)k^2{\cal P}^1h=\kappa {\cal P}^1 \tau\,.
\end{equation}
Rather interestingly, since recalling Eq. (\ref{Grelations}) we know that $a+b=0$, this implies that there are in fact no vector degrees of freedom, and accordingly the stress-energy tensor must have no vectorial part: ${\cal P}^1\tau=0$.

Next let us look at the scalar multiplets. By acting ${\cal P}_s^0$ and ${\cal P}_w^0$ on (\ref{operator-form}) we obtain
\begin{eqnarray}
(a+3d)k^2{\cal P}_s^0h+(c+d)k^2{\sqrt{3}} {\cal P}_{sw}^0h&=&\kappa {\cal P}_s^0\tau\,,  \label{scalar1} \\
(c+d)k^2{\sqrt{3}} {\cal P}_{ws}^0h +(a+2b+2c+d+f)k^2 {\cal P}_w^0h&=&\kappa {\cal P}_w^0\tau\,. \label{scalar2}
\end{eqnarray}
As we see, in principle, the scalar multiplets are coupled.
However, by applying the projector ${\cal P}^0_w$ on equation (\ref{scalar1}) or the projector ${\cal P}^0_s$ on equation (\ref{scalar2}) from the right hand side, one sees that $c+d=0$, in accordance with ({\ref{Grelations}).
The scalars decouple and one can now straightforwardly invert the field equations to obtain the propagators:
\begin{equation}
(a+3d)k^2{\cal P}_s^0h=\kappa {\cal P}_s^0\tau\Rightarrow {\cal P}_s^0h=\kappa {{\cal P}_s^0\over (a+3d)k^2}\tau {\mbox{ and }}
\end{equation}
\begin{equation}
(a+2b+2c+d+4f)k^2 {\cal P}_w^0h = \kappa {\cal P}_w^0\tau\Rightarrow {\cal P}_w^0 h=\kappa {{\cal P}_w^0\over (a+2b+2c+d+f)k^2}\tau\,,
\end{equation}
respectively. The denominator corresponding to the ${\cal P}_w^0$ projector vanishes. So there is no $w$-multiplet, but the $s$-multiplet picks up a nontrivial propagator. 

To finally summarise: 
\begin{equation} \label{result}
\Pi={{\cal P}^2\over ak^2}+{{\cal P}_{s}^0\over (a-3c)k^2}\,.
\end{equation}
We have thus arrived at the main result of \cite{Biswas:2011ar}.


\section{Applications to special cases}
\label{sc}

In this section, we consider the implications of the result (\ref{result}) to some special cases. 

\subsection{General Relativity}

Since we want to recover GR in the infrared, we require from any viable theory that 
\begin{equation}
a(0)=c(0)=-b(0)=-d(0)=1\,,
\end{equation}
corresponding to the GR values. In GR these functions are the same constants for any Fourier mode.  The above condition ensures that as $k^2 \rightarrow 0$, we have only the physical graviton propagator,
\begin{equation}
\lim_{k^2\rightarrow 0} \Pi = ({\cal P}^2/k^2)-({\cal P}^0_s/2k^2) \equiv \Pi_{GR} \,.
\end{equation}
There is a crucial subtlety one should observe here. Although the ${\cal P}^0_s$ residue at $k^2=0$ is negative, that is a benign ghost. In fact, ${\cal P}^0_s$ has precisely the right coefficient to cancel the unphysical longitudinal degrees of freedom in the spin-2
part ~\cite{VanNieuwenhuizen:1973fi}. 

\subsection{Gauss-Bonnet gravity}

Let us consider the theory ${\cal L}=R+\alpha(\Box) G$, where $G$ is the Gauss-Bonnet invariant $G=R^2-4R_{\mu\nu}R^{\mu\nu}+R_{\mu\nu\rho\sigma}R^{\mu\nu\rho\sigma}$ and the function $\alpha(\Box)$ in the simplest case can be just a constant coefficient. As is well known, in four dimensions the Gauss-Bonnet term is a topological invariant that does not contribute to the gravitational field equations. Therefore it is not a surprise that it doesn't introduce any modifications to the propagator either. In Eq.(\ref{sq}) we have now $F_1=\alpha$, $F_3=-4\alpha$ and $F_{10}=\alpha$.  Regardless of $\alpha$ we then see from Eqs.(\ref{coef}) that $a=c=-b=-d=1$, and thus the properties of the theory are identical to GR.

\subsection{{\cal L}(R) gravity}

The ${\cal L}(R)$ gravity is a popular subject of study. For our purposes here, it is enough to consider the expansion of the lagrangian around flat space,
\begin{equation}
{\cal L}(R)={\cal L}(0) + {\cal L}'(0)R + \frac{1}{2}{\cal L}''(0)R^2 + \cdots\,.
\end{equation}
The first term one identifies with the cosmological constant, ${\cal L}(0)=-2\kappa^{-1}\Lambda$, and the second term should reduce to the Einstein-Hilbert term in a viable theory, ${\cal L}'(0)=1$. The relevant modification of the theory is then given by the
quadratic part. Since only $F_1$ is now nonzero in (\ref{sq}), we readily see from (\ref{coef}) that then $a=-b=1$, $c=-d=1-{\cal L}''(0)\Box$ and $f=-{\cal L}''(0)\Box$. The propagator is thus
\begin{equation}
\Pi = \frac{{\cal P}^2}{k^2} - \frac{{\cal P}^0_s}{2k^2(1+3{\cal L}''(0)k^2) }\,.
\end{equation}
The scalar part of the propagator is modified. Since these theories are a specific class of scalar-tensor theories, we expect an extra scalar degree of freedom. Its appearance can be made transparent by rewriting the above result as
\begin{equation}
\Pi = \Pi_{GR} + \frac{1}{2}\frac{{\cal P}^0_s}{k^2+m^2}\,, \quad m^2 = \frac{1}{3{\cal L}''(0)}\,.
\end{equation}
Indeed, the ${\cal L}(R)$ correction entails an additional spin-0 particle which is nontachyonic as long as\footnote{For an alternative derivation and generalisation of this stability condition, see \cite{Amendola:2010bk}.} ${\cal L}''(0)>0$. One also sees that though these theories are classically viable, they cannot improve the ultraviolet properties of GR since the graviton propagator retains its form.  

\subsection{Conformally invariant gravity}

As an example of a ghastly theory, let us consider the Weyl squared gravity. The Weyl tensor is defined as
\begin{equation}
C_{\mu\nu\rho\sigma} = R_{\mu\nu\rho\sigma} + \frac{R}{6}\left( g_{\mu\rho}g_{\nu\sigma} - g_{\mu\sigma}g_{\nu\rho}\right) - \frac{1}{2}\left( g_{\mu\rho}R_{\nu\sigma} - g_{\mu\sigma}R_{\nu\rho}-
g_{\nu\rho}R_{\mu\sigma} + g_{\nu\sigma}R_{\mu\rho}\right)\,. \nonumber
\end{equation}
The theory is then specified by the conformally invariant Weyl-squared term, ${\cal L}=R-\frac{1}{m^2}C^2$, where $m$ is the mass scale at which the correction becomes relevant.  It is straightforward to compute that
\begin{equation}
C^2= R_{\mu\nu\rho\sigma} R^{\mu\nu\rho\sigma} -2R_{\mu\nu}R^{\mu\nu}+\frac{1}{3}R^2\,,
\end{equation}
from which we quickly infer, using again Eqs.({\ref{sq},\ref{coef}) that now $a=-b=1-(k/m)^2$, $c=-d=1-(k/m)^2/3$ and $f=-2(k/m)^2/3$. We obtain the propagator with a double pole for the graviton:
\begin{equation}
\Pi =  \frac{{\cal P}^2}{k^2\left( 1-(k/m)^2\right)} - \frac{{\cal P}^0_s}{2k^2} =  \Pi_{GR} - \frac{{\cal P}^2}{k^2+m^2}\,.
\end{equation} 
From the latter form of the propagator it is obvious that the theory contains an extra spin-2 degree of freedom with respect to GR. Moreover, the extra contribution always comes with the wrong sign: this is the Weyl ghost\footnote{However perhaps the negative norm states can be consistently projected out of the Hilbert space \cite{Mannheim:2011ds}.}.

\subsection{Asymptotically free gravity}

Finally, we show how the ultraviolet properties of GR are improved via a nonlocal extension. Just for simplicity, let us restrict to the special class of theories with $f=0$. From (\ref{Grelations}) we then see that
$a=c=-b=-d$. This means that we are essentially left with just a single free function
\begin{equation}
a(\Box)= 1- \frac{1}{2}F_3 (\Box) \Box - 2 F_{10} (\Box) \Box\,.
\end{equation}
We obtain a very  simple expression for the propagator:
\begin{equation}
\Pi={1\over k^2a(-k^2)}\left(  {\cal P}^2-\frac{1}{2}{\cal P}_{s}^0\right)=\frac{1}{a(-k^2)}\Pi_{GR}\,.
\end{equation}
Thus, the GR propagator is now modulated by the $k$-dependent function $a(\Box)$. We now realise that as long as $a(\Box)$ has no zeroes, these theories contain no new states as compared to GR, and only modify the physical graviton propagator. Polynomial functions would correspond to higher - but finite - order gravity, and would inevitably result in new pathological states. This can be avoided in nonlocal, i.e. infinite order higher derivative theories.  Furthermore, by choosing $a(k^2)$ to be a suitable entire function we can indeed tame the behavior of the ultraviolet gravitons. A simple example can be provided by $a=\exp{(k/M)^2}$, where $M$ is a mass scale at which the nonlocal modifications become important. The integrals over the propagator quickly tend to zero at high momenta $k>M$ and we expect finite results from physical calculations (note though that light-like momenta do not receive damping).
. 

\section{Conclusions}

Having derived the main result (\ref{result}), we considered its implications in some special cases. We readily reproduced the known results: while GR and Gauss-Bonnet theory share the same field content, $f(R)$ gravity has an extra healthy scalar and Weyl gravity an extra pathological spin-2 field. New classes of nonlocal theories were found, that are both unitary and devoid of singularities. The ongoing further work includes the generalisation of the result to curved backgrounds.


\setcounter{equation}{0}
\appendix
\acknowledgements{TK would like to thank  Erik Gerwick and Alex Koshelev for their contributions to these calculations and
Danielle Wills, Sergey Vernov and Nicola Tamanini for insightful discussions. TB is supported by the Louisiana Board of Regents, TK by the Research Council of Norway and AM by the STFC grant ST/J000418/1.} 


\end{document}